\documentclass{sig-alternate}

\usepackage[english]{babel}
\usepackage{cite}
\usepackage{graphicx}
\usepackage{url}
\usepackage{xspace}
\usepackage{balance}
\usepackage{color}

\usepackage{enumitem}
\newlength{\descindent}
\newcommand{\abqm}{\textsc{ABQM}\xspace}

\newcounter{Examplecount}
\newtheorem{question}{RQ}

\begin{document}

%
\conferenceinfo{WoSQ'11,} {September 4, 2011, Szeged, Hungary.} 
\CopyrightYear{2011} 
\crdata{978-1-4503-0851-9/11/09} 
\clubpenalty=10000 
\widowpenalty = 10000



\title{The Use of Application Scanners\\ in Software Product Quality Assessment\titlenote{This 
work has partially been supported by the German Federal Ministry 
of Education and Research (BMBF) in the project QuaMoCo (01 IS 
08023B).}}

\numberofauthors{1}

\author{
\alignauthor Stefan Wagner\\  
  \affaddr{Institute of Software Technology\\
  		University of Stuttgart\\
                 Stuttgart, Germany}\\
  \email{stefan.wagner@informatik.uni-stuttgart.de}
}

\maketitle

\begin{abstract}
Software development needs continuous quality control for a timely detection
and removal of quality problems. This includes frequent quality assessments,
which need to be automated as far as possible to be feasible. One way of
automation in assessing the security of software are application scanners
that test an executing software for vulnerabilities. At present, common quality assessments
do not integrate such scanners for giving an overall quality statement. This
paper presents an integration of application scanners into a general quality
assessment method based on explicit quality models and Bayesian nets. 
Its applicability and the detection capabilities of common scanners are
investigated in a case study with two open-source web shops.
\end{abstract}




\section{Introduction}
\label{sec:intro}

Continuous quality control means to assess and improve software quality 
almost continuously, i.e., on an hourly or daily basis.  This allows the developers
to detect quality defects and to remove them early after their introduction into the system,
which avoids a general quality decay and far higher costs in later phases in the
software's life cycle. These benefits, however, come at the cost that quality assessments
need to be done often and hence are elaborate. Therefore, automation and good tool
support is necessary to employ continuous quality control in practice 
\cite{2008_deissenboeckf_quality_control}.

\subsection{Problem Statement}

Software product quality assessments need to cover a large variety of topics including
security. The assessment of product security is -- as all
quality analyses -- elaborate. Hence, also for security, automation is necessary
for practical adoption. In quality assessments automation relies to a large degree on 
automated static analysis. Static analysis, however, can only assess security partial.
Dynamic analyses are needed to complement the static ones. Most existing automatic 
dynamic analyses for security are not integrated into product quality assessment methods.
Instead, dynamic analysis tools, mostly so-called application scanners, are used solely for 
analysing the security of networks and hosts.

\subsection{Research Objective}

Similar to static analysis tools, there is a a plethora of tools
for automatic dynamic security analysis. Especially application scanners are
available commercially as well as open source.
Those tools scan the executing application automatically for vulnerabilities
and hence are a promising addition to static analysis.
Our overall objective is to investigate the available tools and the kinds of vulnerabilities
they detect to define how these tools should be integrated in a general quality
assessment.

\subsection{Contribution}

We employ an
existing quality assessment method based on explicit quality models and Bayesian
nets and extend it by defining how application scanners can be used in the assessment.
This extended method is performed using three well-known open source application scanners
(w3af, Wapiti, and Grendel Scan) on two open source web shops (PHP Shop and Zen Cart).
We show the principal applicability of the method to these real-world applications and
also find first indications that the scanners find different vulnerabilities and, hence, should
be used in combination.
Therefore, this paper is only a first step in the direction of the research objective. 

\subsection{Context}

The approach is applicable in principle to any kind of software. Most application
scanners focus on web applications at present. The used application scanners and study object
are open source but in use in commercial contexts.

\subsection{Outline}
We start by introducing application security scanners and especially the used scanners
in section~\ref{sec:scanner}. We then explain the quality assessment method based on
an explicit quality model and Bayesian nets that we will use as an example method into
which we integrate the scanner results in section~\ref{sec:qam}. In section~\label{sec:case},
we describe the design and results of the case study. Finally, we compare our results
with related work (section~\ref{sec:relwork}) and give final conclusions (section~\ref{sec:conclusions}).

\section{Application Security Scanners}
\label{sec:scanner}

We first give a general introduction into what application security scanners are and present
three scanners that are also used in the case study in Section \ref{sec:case}.

\subsection{General}

In general, an application scanner is a software that performs automatic penetration testing.
Most scanners use a set of common patterns of inputs that they send to the application and
decide, based on the output, whether there is a vulnerability that might be exploited. In
addition, they have many possibilities to configure the penetration tests so that they
fit to the system under analysis. Most
application scanners concentrate on web applications as these are most exposed to attacks.
Black et al.\ define in \cite{black08} a \emph{Web application security scanner} as
an ``automated program that searches for software security 
vulnerabilities within web applications''.

There are several groups that work on specific application scanners (e.g., \cite{huang04,kals06,balzarotti07}) in order to either find
new vulnerabilities or improve the detection of vulnerabilities. There are also specialised
tools that dynamically and (partly) statically detect specific vulnerabilities \cite{balzarotti08}. However, these different tools
have not been compared and analysed w.r.t.\ their usage in product quality assessment.
We discuss three common open-source scanners in the following.

\subsection{w3af}

The \emph{Web Application Attack and Audit Framework} (w3af) provides a framework as well
as a complete graphical and command-line interface to run application scans and
view results. The framework provides simple wrappers for HTTP communication, web services, 
sessions, and HTML parsing. It also contains many plugins that implement scanning and testing
an application. It is written in Python and is available at
\url{http://w3af.sourceforge.net/}.

\subsection{Wapiti}

The \emph{Web application vulnerability scanner / security auditor} (Wapiti) is a command-line
tool that scans the web pages of an application and identifies scripts and forms to inject data.
Using these scripts and forms it acts like a fuzzer and injects payloads to see if a script is vulnerable.
Wapiti is developed in Python. It is
available at \url{http://wapiti.sourceforge.net/}.

\subsection{Grendel-Scan}

Grendel-Scan is a web application security testing tool that also provides a
graphical user interface. It contains an automatic application
scanner that detects common web application vulnerabilities. It is written in Java and  is available at
\url{http://www.grendel-scan.com/}.

\section{Quality Assessment Method}
\label{sec:qam}

Quality assessment is the part in quality control that compares the actual state of an application
with its requirements. It evaluates if and how well the software
fits to what was intended. 
There are various ways to perform this assessment and the major difficulty lies in
combining the various quality assurance results and measures to a common quality
statement. In the project Quamoco\footnote{\url{http://www.quamoco.de/}}, we developed such a quality assessment method. One
specific instance of this method uses Bayesian nets to describe the uncertainties in the 
results and measures as well as to calculate a quality statement.
In the following,
we propose how application scanners can be integrated in the method for a substantial
security assessment.

\subsection{Quamoco}

In the project Quamoco, we develop a quality model
with a corresponding quality assessment method 
that has the aim to facilitate continuous improvement based on objective, quantitative
feedback \cite{lampasona09}. It has its origins in the Quality Improvement Paradigm~\cite{basili1988tame} and the 
Goal/Question/Metric (GQM) approach~\cite{basili94}. We built one specific instance using Bayesian nets
as a means for analysing assessment results \cite{wagner:ist10} that was
specifically aimed at using activity-based quality models \cite{Deissenboeck2007ActivityBased}. 

We give a brief overview on the quality models developed in Quamoco and 
describe the assessment method from \cite{wagner:ist10} adapted to
the Quamoco quality models.

\subsection{Quamoco Quality Models}

In general,
there are two main uses of quality models in a software project: (1) as a basis
for defining quality requirements and (2) for defining quality assurance
techniques and measurements for the quality requirements. 
The quality models developed in Quamoco advance existing quality 
models as they combine the practically shown advantages of different models 
\cite{Deissenboeck2007ActivityBased,Winter07acomprehensive,ploesch08,grossmann09}. 
The idea is to use not only high-level ``-ilities'' for defining quality but instead to break
it down into detailed factors and their influence on \emph{quality attributes}.
The quality attributes we use in this paper are the activities performed on and 
with the system, which are derived from activity-based quality models \cite{Deissenboeck2007ActivityBased}.
In the area of security, we use a hierarchy of attacks as activities \cite{wagner:qaw09}; in this
case activities that should be prevented.

We developed an explicit meta-model in Quamoco that defines the
 quality model elements and their relationships. Five
elements of the meta-model are most important in the context of this paper: \emph{entity}, 
\emph{property}, \emph{measure}, \emph{impact}, and \emph{activity}.  An \emph{entity}
can be any thing, animate or inanimate, that has an influence on the software's 
quality, e.g., the source code of a PHP function or an HTML form. These 
entities are characterised by properties such as \emph{structuredness} 
or \emph{conformity}. The combination of an entity and a property is
called a \emph{factor}. 
These factors are measurable
either by automatic measurement or by manual review. 
This is specified in the \emph{measures} for a factor.

Entities as well as activities are organised in hierarchies.
An influence of a factor is specified by an \emph{impact}. We concentrate
on the influences on attack activities, for example, \emph{SQL injection} or \emph{password brute forcing}.
The impact on an activity can be positive or negative. 

\begin{figure}[htb]
\centering
\includegraphics[width=0.4\textwidth]{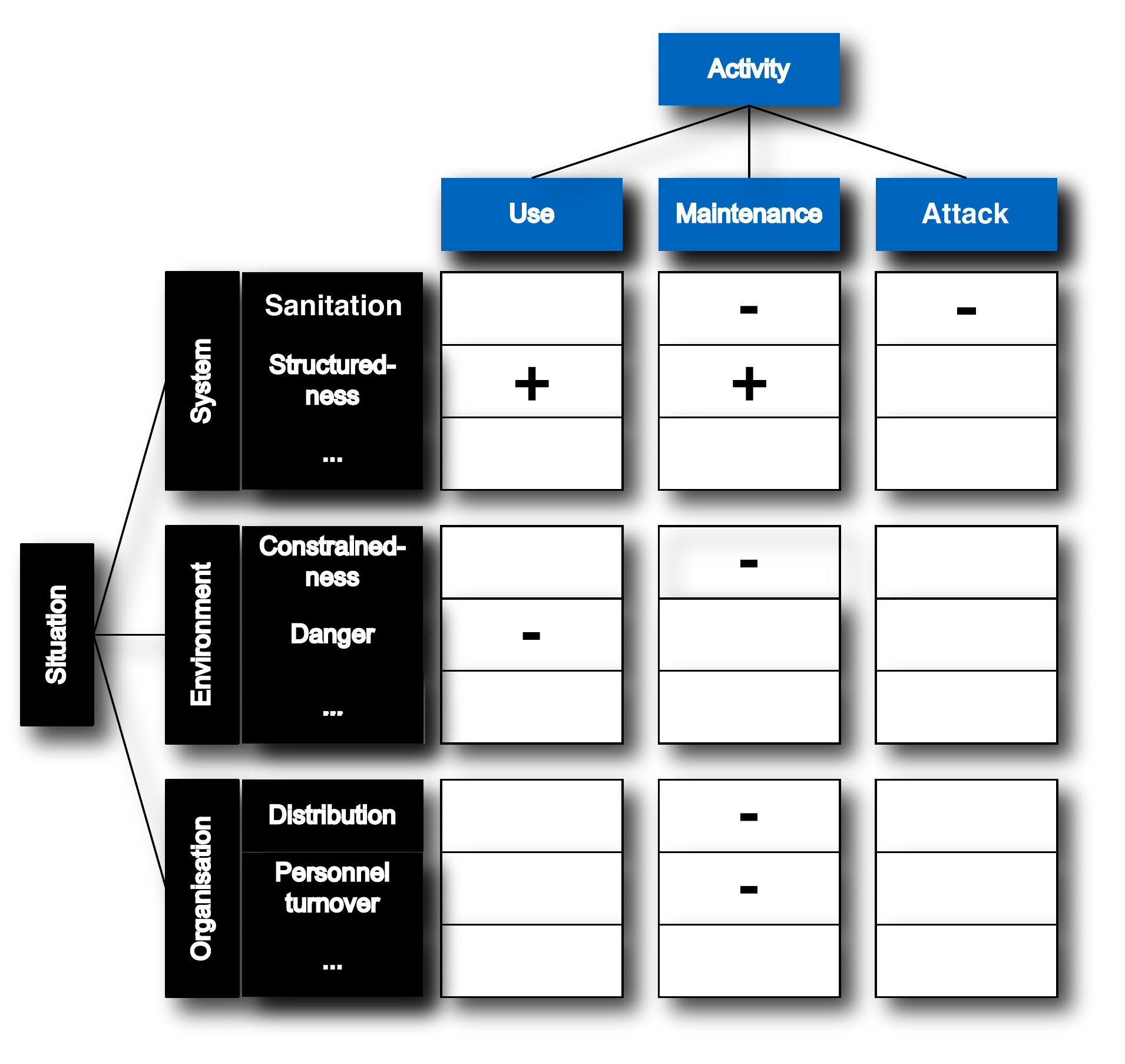}
\caption{High-level view on an activity-based quality model as a matrix 
(based on \cite{wagner:ist10})}
\label{fig:quality_matrix}
\end{figure}

The two hierarchies, the factor tree and the
activity tree, together with the impacts of the facts on the activities can
be visualised as a matrix (Figure \ref{fig:quality_matrix}). The 
factor tree is shown
on the left, the activity tree on the top. The impacts are depicted by entries
in the matrix where a ``+'' denotes a positive and a ``--''
a negative impact.
The associations between factors in the factor tree denote a ``kind-of'' 
relationship.

\subsection{Bayesian Nets}

Bayesian nets are a means for modelling uncertain relationships that can
be simulated and that predicts probable outcomes. They are a
modelling technique that can represent causal relationships based on
Bayesian inference. They are represented as a directed acyclic graph 
with nodes for uncertain variables and edges for directed relationships
between the variables. This graph models all the relationships abstractly.

For each node or variable there is a corresponding \emph{node probability 
table} (NPT). These tables define the relationships and the uncertainty in
these variables.  For each state of each variable, the probability that the 
variable is in this state is specified. If
there are parent nodes, i.e., a node that influences the current node,
these probabilities are defined in dependence on the states of these 
parents. 
A complete Bayesian net allows the forward and backward calculation
for different scenarios based on observations or desired outcomes.

\subsection{Steps}

The assessment method \cite{wagner:ist10} consists of four steps for building a Bayesian net
derived systematically from a Quamoco quality
model. The resulting Bayesian net contains three
types of nodes:
\begin{itemize}
\item \emph{Activity nodes} that represent activities from the quality model
\item \emph{Factor nodes} that represent factors from the quality model
\item \emph{Measure nodes} that represent indicators for activities or factors
\end{itemize}
We need four steps to derive these nodes from the information of the
quality model. 

\begin{enumerate}
\item We identify the relevant activities with measures
based on the assessment goal. We use GQM \cite{basili94} to structure
 that derivation. We first define the
assessment  \emph{goal}, for example, optimisation of security assurance,
which leads to relevant activities, such as \emph{attack}. This is refined by
stating \emph{questions} that need to be answered to reach that goal.
\item Influences by
sub-activities and factors are identified. This step is repeated recursively
for sub-activities. The resulting factors together with their impacts are modelled.
\item Suitable measures for the factors are added. 
\item The node probability
tables (NPT) are defined to reflect the quantitative relationships. This includes 
defining node states as well as filling the NPT for each node. The
activity and factor nodes are usually modelled as \emph{ranked} nodes, i.e., in an
ordinal scale. Having that,
the Bayesian net can be used for simulation by setting values for any
of the nodes. 
\end{enumerate}

The definition of NPTs is the most complicated part in building Bayesian nets.
The approach by Fenton, Neil and Galan Caballero \cite{fenton07a} simplifies
that by approximating the specific values in an NPT by general
distributions or expressions. They formalise the behaviour observed with experts that have to
estimate NPTs, who usually estimate the central tendency or some extreme values
based on the influencing nodes. The remaining cells of the table are then filled
accordingly.  For example, it renders it possible to model the NPT
of a node by a weighted mean over the influencing nodes.

In general, the NPTs of the measure nodes
are defined using either common industry distributions or information from
company-internal measurements. The influence of the activity or factor node
it belongs to can be modelled in at least two ways: (1) partitioned expressions
and (2) arithmetic expressions. The latter describes a direct arithmetical
relationship from the level in the activity or factor node to the measure. Using
a partitioned expression, the additional uncertainty can be expressed by
defining probability distributions for each level of the activity or factor node.

\subsection{Integration of Application Scanners}

Application scanners provide findings of probable vulnerabilities for the analysed software.
We can use them as measures for factors. Hence, the integration
of application scanners affects the steps 3 and 4 of the assessment method. We define
measure nodes that correspond to scanner findings. All scanners classify
the found vulnerabilities into different types. Each vulnerability type forms a measure.
These measures are matched to existing factors or new factors are generated enriching
the knowledge about how to develop secure software applications.

For example, an application scanner might detect buffer overflows if the software is
configured to return error pages. The assessment method user would create a measure
node \emph{Buffer Overflow Error Page} that represents the findings of the scanner. The
quality model already contains a factor \emph{Confinement of Buffer}, which
specifies that the limits of buffers are respected. This factor is represented in the
Bayesian net as a factor node and the assessor adds an influence to the measure node.

The factors that are measured by application scanners can have an impact on a very 
specific attack or in general ease attacking. This is reflected by the hierarchy level of
the attack that has the impact. A general impact goes to a more generic attack in the activity
hierarchy. For the example of the buffer overflow, the impact might be on the attack
\emph{Forced Integer Overflow} that represents the setting of a controllable integer
value to an unexpected value.

For measures from static analysis, we calculate densities to reflect
how large the problems are in relation to the software size. As each found vulnerability
can potentially corrupt the complete application, we use a simpler yes/no voting. If there
is at least one vulnerability of a type, the measure has the value \emph{yes}. For example,
if the scanner detects at least one buffer overflow error page, the assessor sets the observation of the
measure node to \emph{yes}. The NPT in the measure node is modelled by a partitioned
expression. In the buffer overflow example, if \emph{Confinement of Buffer} is in the state
\emph{high}, \emph{Buffer Overflow Error Page} is in the state \emph{no} and vice-versa. The
expression should also add an uncertainty range depending on how well the measure
indicates the factor.

If we employ more than one scanner, we can run into the problem that the scanners
do not agree on the detection of specific vulnerabilities. We prefer a pessimistic assessment
-- possibly worse than it actually is -- and hence vote \emph{yes} if at least one scanner
reports a vulnerability.

\section{Case Study}
\label{sec:case}

The case study shows the applicability of the method and to
a smaller degree the detection capabilities of application scanners.
We define the study design, describe the used study
objects, and show and discuss the results.

\subsection{Study Design \& Procedure}

The aim of this case study is  a proof-of-concept that analyses the method's applicability
to real-world software. In particular, we are interested in the effort needed to
incorporate and use the scanners as well as if they give useful results. Furthermore,
the execution time for analysis should be short enough to be able to run the scanners
often, e.g., on a daily basis. This leads to our first research question:
\begin{question}
Is the assessment method applicable to realistic software systems?
\end{question}

Moreover, we investigate if common scanners are
comparable in terms of the vulnerabilities they detect. The
experience with static analysis has shown that different tools
detect partly different classes of defects. If this is not the case, we could
resort to just one tool in quality control, which would reduce our
effort considerably. Hence, our second research question
asks for the differences in vulnerability detection:
\begin{question}
Are there differences between the detection capabilities
of different application scanners?
\end{question}

We analyse both questions by applying 3 widely known open-source
application scanners (see Section \ref{sec:scanner}) to 2 open-source
web shops. We install both web shops with their
standard installation and run each scanner
on each web shop. The scanners are configured
to reasonable settings w.r.t.\ the study objects. For example, attacks
specifically for Microsoft SQL Server make no sense as a MySQL
database system is used by the study objects.

The vulnerabilities found by all scanners are partitioned into classes
that stem from the types of vulnerabilities found by the scanners. The
classes are used in the quality assessment method to make a quality statement
about the study objects. The Bayesian net for that is built using the tool AgenaRisk. 
We analyse this application of the method
qualitatively to answer RQ 1. Then we compare the results of all three scanners
separately and  compare their findings for answering RQ 2. The comparison analyses
to what degree there are overlaps in the found classes of vulnerabilities. The 
vulnerabilities are not checked for false positives.

\subsection{Study Objects}

The study
objects are two different web shops, one -- Zen Cart -- a large application, which is also the
most popular of this kind on sourceforge. The other application -- PHP Shop -- is
simple and small in comparison to Zen Cart. Hence, in the case selection, triangulation
 is used as far as possible. Detailed descriptive
information about both study objects is given in Table \ref{tab:objects}.

\begin{table}[htp]
\caption{Information about the study objects}
\begin{center}
\begin{tabular}{lrr}
\hline
 & PHP Shop & Zen Cart\\
\hline
Language & PHP & PHP/Perl \\
1.~Release & 1999 & 2004 \\
Database    & MySQL & MySQL \\
Used Version & 0.8.1 & 1.3.8 \\
SLOC & 8,052 &  73,001\\
Downloads & 53,000 & 625,000\\
\hline
\end{tabular}
\end{center}
\label{tab:objects}
\vspace{-2em}
\end{table}%

Both applications were installed in the standard Apache web server 
available in Mac OS X and connected to a local MySQL installation as
database management system. As far as possible all configuration were
left with the default values.

\subsection{Results}

As a result for the applicability of the approach, we describe the concrete
application together with our experiences.
We start with the first step of our assessment approach and
 identify the relevant activities and corresponding measures. We
analyse security, in particular the risk of vulnerabilities in the system. 
The risk can be the basis for deciding whether further security improvements
needs to be employed.  Therefore, the \textbf{goal} is ``Planning of further security
improvements''. For security improvements, attacks on the system need to
be confounded. Hence, the activity \emph{Attack} needs to be analysed.
We derive the \textbf{question} ``How many 
vulnerabilities are there in relation to the software size?''. For the security
improvement planning, it is
not only important how many vulnerabilities there are but also whether this
number is in a reasonable relation to the system size. It might be economically
inadvisable to invest in removing all vulnerabilities.  The corresponding \textbf{metric} 
\emph{vulnerability density} that measures the number of vulnerabilities by source
code size in KSLOC can be directly derived from the question.

\begin{figure*}[htbp]
\begin{center}
\includegraphics[width=.8\textwidth]{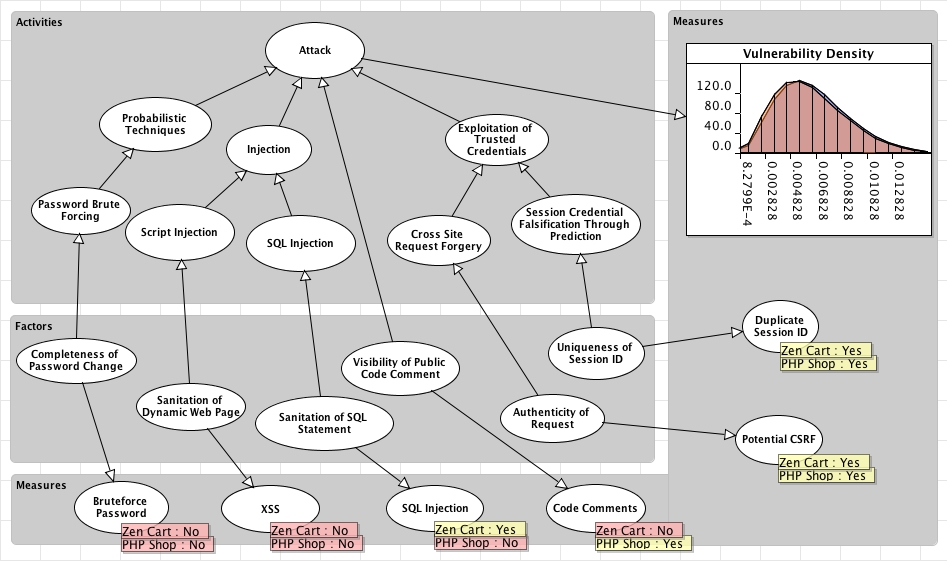}
\caption{The Bayesian net for the analysis from AgenaRisk}
\label{fig:bn}
\end{center}
\end{figure*}

In the second step of the assessment method, we build the
Bayesian net. The selection of the nodes in the study is driven by the detection
possibilities of the used scanners. There is the top-level activity \emph{Attack}
that we measure by the above derived \emph{vulnerability density}. It has 
a direct impact from \emph{Visibility of Public Code Comment} that describes that it
is easier to attack if there are code comments visible to the public. Then, we break \emph{Attack} 
down to \emph{Probabilistic Techniques}, \emph{Injection}, and \emph{Exploitation of
Trusted Credentials}. These are further refined into 
\emph{Password Brute Forcing}, \emph{Script Injection}, \emph{SQL Injection}, 
\emph{Cross site Request Forgery}, and \emph{Session Credential Falsification Through
Prediction}. Figure \ref{fig:bn} shows in the top left the activity tree as represented in
the Bayesian Network.

We include 6 impacts on these activities. The impacts are chosen so that their
corresponding factors can be measured by the investigated application scanners. 
The factors used are:
\begin{itemize}
\item
\emph{Completeness of Password Change}: Any implementation of changing user
passwords is also responsible for the quality of that password to avoid password
brute forcing. If such a check is missing, we consider the implementation to be incomplete.

\item
\emph{Sanitation of Dynamic Web Page}: If a web application does not sufficiently 
sanitise the data it is using in output, arbitrary
content, including scripts, can be included by attackers.

\item
\emph{Sanitation of SQL Statement}: Analogously to dynamic web pages, the used
SQL statements need to be sanitised to avoid unwanted changes or reads to the 
database.

\item
\emph{Visibility of Public Code Comment}:
Comments in HTML or Java Script code visible to the public may give attackers
information they can exploit.

\item
\emph{Authenticity of Request}:
The application needs to be able to undoubtedly decide on the authenticity of a
request. If this is not the case, Cross Site Request Forgery is possible.

\item
\emph{Uniqueness of Session ID}: Each session needs a unique ID that cannot
easily be guessed. Otherwise an attacker may predict an ID and gain access to
the application

\end{itemize}

In the fourth step of the approach, measures are defined for all impacts. The measures
here are derived from the vulnerabilities identifiable by the scanners and are attributable to
the application -- as opposed to the environment.
The final topology of the Bayesian net is shown in Figure~\ref{fig:bn}. Overall,
building the Bayesian net took less than a day.

The execution of the scanners took between several minutes (PHP Shop) and
several hours (Zen Cart) on a current MacBook Pro that runs both the web server
and the scanners. The found vulnerabilities are shown in Table \ref{tab:vulnerabilities}.
Waipiti did not find any vulnerabilities in both cases. We analysed its execution
in detail to avoid any misconfigurations, but it seems that it is not able to detect
problems in the analysed software. 
Most vulnerabilities were detected by Grendel Scan, 3 vulnerabilities were reported
by w3af.

\begin{table}[htp]
\caption{The vulnerabilities found in the scans. The characters A--C denote which
scanner found the vulnerability: A=w3af, B=Wapiti, C=Grendel Scan.}
\begin{center}
\begin{tabular}{lcc}
\hline
Vulnerability & PHP Shop & Zen Cart\\
\hline
Duplicate Session ID & C & C \\
Potential CSRF           & A, C & C\\
SQL Injection              &     & A, C \\
Code Comments        & C & C \\
Unidentified Vuln.      & A & \\
Input/Output Flows     & C & \\
\hline
\end{tabular}
\end{center}
\label{tab:vulnerabilities}
\vspace{-2em}
\end{table}%

This information was then used in the Bayesian net to assess the quality of the two 
applications. Two vulnerability classes from Table~\ref{tab:vulnerabilities}, the 
input/output flows and unidentified vulnerabilities, were not further
used because they cannot be attributed to a specific product entity. The predicted
vulnerability densities (vulnerabilities/KSLOC) of both applications are very close. The net
calculated a mean of 0.0064 for PHP Shop (standard deviation 0.0028) and a mean
of 0.0066 for Zen Cart (standard deviation 0.0028).

\subsection{Discussion \& Threats to Validity}

The assessment method including application
scanners is applicable to the real-world systems we analysed. Yet, the results are
close for both systems and the correctness of the results cannot be validated as we
have no data about the real vulnerability density.
Nevertheless, the effort for performing the assessment is reasonable. The setup
of the scanners and a corresponding test environment is more demanding than
the subsequent analysis using the Bayesian net. Altogether it took only a few days to
set up the analysis. Also the time needed for running
the scanners is promising and allows them to be included in continuous quality
control.

An important decision in modelling application security is the border between the application
that is analysed and its environment. For example, is the application responsible
for passwords that are not prone to brute force attacks? In the case study we made
subjective choices and for the example of passwords specified that the application
has a partial responsibility.

The answer to RQ 2 is more clear as the found vulnerabilities differ between
the used scanners. Wapiti did not find a single vulnerability, Grendel Scan found
8 and w3af found 3 vulnerabilities. For potential cross site request forgery in PHP Shop and
SQL injection in Zen Cart both scanners had findings. However, only Grendel Scan
detected a potential CSRF in Zen Cart. All these differences indicate that there are
significant differences between the detection capabilities of different scanners. 

As this is only a first, explorative study on the use of application scanners in quality
assessment, there are various threats to the validity of the results. The internal
validity is threatened because there were several subjective decisions in building
the Bayesian net. We mitigated this threat by using comparable decisions as for
static analysis. Furthermore, we did not check whether the found vulnerabilities
are actual problems in the software. This especially affects RQ 2, because the
results might be misleading.
The external validity is also limited as we only analysed two applications and three
scanners, which are all open source. For more reliable results, especially for the
detection capabilities, we need to run larger studies that also involve commercial
applications and scanners.

\newpage
\section{Related Work}
\label{sec:relwork}

We discuss quality models, guidelines and measures and especially several
security assessment approaches.

There is a wide variety of quality models. Deissenboeck et al.~\cite{2009_deissenboeckf_quality_models} differentiate
between quality definition models and quality assessment models. The former is a specification of what
constitutes quality in a software system, the latter describes how a software system's quality can be assessed according
to specific rules. In the area of software security, security pattern collections are an example of quality definition
models, e.g., \cite{hafiz07}.

\emph{Quality definition models} are either general but too abstract for a concrete use in assessing software quality (e.g., ISO 9126) or
specialised for a specific quality attribute and hence difficult to integrate into general quality assessments \cite{Deissenboeck2007ActivityBased}.  
In~\cite{Deissenboeck2007ActivityBased}, Deissenboeck et al.\ propose a quality model (\abqm) that tackles this problem by breaking quality
attributes into entities, their properties, and their influence on activities. 
In~\cite{wagner:qaw09} we used the ABQM approach for modelling security but with a focus on security requirements.

\emph{Quality guidelines} are developed by various companies and organisations and usually include technical aspects that have to be taken into account. For example, the Common Criteria catalog (CC)~\cite{CommonCriteria} and the German BSI \emph{IT-Grundschutz} Manual~\cite{BSI} describe security requirements. Usually guidelines do not give rationales ~\cite{Deissenboeck2007ActivityBased}. Hence, they do not guide through a structured process, are often read once and followed in a sporadic manner only~\cite{Broy2006Demystifying}. Furthermore, it is often not checked whether guidelines are followed or not~\cite{Deissenboeck2007ActivityBased}.

Common \emph{metric-based/stochastic approaches} describe quality by measurable concepts that imply strong assumptions. While for some quality attributes, those assumptions are stable for others, such as security, the assumptions are changing fast~\cite{Aime2008SecMet}. Due to their single-value representation, metrics often do not explain how system properties influence the quality related activities that are performed with the system \cite{Deissenboeck2007ActivityBased}. 
Hence, metrics are not well established for security~\cite{Atzeni2006Why} and unstable due to fast variation of the security underlying ``physics'' (i.e., the IT system)~\cite{Aime2008SecMet}. 

Artsiom et al.~\cite{artsiom08} propose an security assessment method that has similarities to
the method in this paper. It also defines metrics and aggregates them to quality attributes. 
This method, however, uses ``-ilities'' similar to ISO 9126 that have several well-known problems. Moreover,
they concentrate on the architecture of the software (white-box view) whereas this paper focuses
on testing by application scanners (black-box view). 

Frigault et al.\ \cite{Frigault2008DBN} use Dynamic Bayesian Networks to investigate the security of networked systems. 
Their focus is more on the combined effects of different vulnerabilities as opposed to a complete
quality statement for the system incorporating scan results.

There are several so-called scoring systems that evaluate vulnerabilities in applications. The most advanced scoring system
is the \emph{Common Vulnerability Scoring System} (CVSS) \cite{CVSSPaper,CVSS}. It provides a set of metrics and 
corresponding equations that combine these metrics with weights to provide a score for a vulnerability. It considers the constraints
as well as the impacts of a vulnerability, but does describe how to find vulnerabilities and how to relate the results to a general quality assessment.

Recently the \emph{Common Weakness Scoring System} (CWSS)~\cite{cwss} was released that analyses
weaknesses in a software system and assigns scores for it for prioritising the weaknesses. One part of the scoring is the technical impact.
Hence, there are similarities to the Quamoco quality model, which we should exploit in the future. By itself, the CWSS describes not how it
fits into an overall quality assessment.

\emph{The Open Web Application Security Project} (OWASP) is a non-commercial initiative to develop guidelines and standards for the security of
web applications. Their \emph{OWASP Application Security Verification Standard 2009} (ASVS) \cite{asvs09} defines 4 \emph{security verification levels}
that describe what has to be done to provide appropriate security for an application. The developer of an application decides on its criticality and the standards
gives the corresponding verification requirements that have to be met. This ranges from mostly automatic analysis to complete manual code reviews. A
level is reached if all these requirements are checked. It does not contain more fine-grained evaluations and it is also not set into the context of a general
quality assessment.

\section{Conclusions}
\label{sec:conclusions}

We summarise the contribution of this paper and discuss
directions for future research.

\subsection{Summary}

Application security scanners, as employed in the area of web applications,
are one promising possibility to automate the assessment of software application
security. This automation could then be used in product quality assessments in the 
context of continuous quality control. However, the usage of application scanners in
this kind of quality assessment has not been investigated so far.

We provide a first step to incorporate application scanners into quality assessment by
extending an existing method based on explicit quality models and Bayesian nets.
In the Quamoco quality models, measures are defined to make use of the scanning results. It is also defined
how these results can be further used for a general quality statement.

We show in a case study how three open source application scanners can be used in the
quality assessment of open source web shop applications. We found that the method is
applicable in principle and that the detection capabilities of the scanners differ.
Moreover, the needed time for performing the scans is promising for their inclusion
into continuous quality control.

\subsection{Future Work}

A threat for the case study is that only three scanners are used. We plan to evaluate more
application scanners, especially tools developed commercially. For a more reliable result
we also plan to investigate further cases involving software developed in industry for which
we also analyse the false positive rate of scanners.
 
The assessment method will be extended to be able to handle false positives
explicitly. The found differences between scanners might also be an indication for false
positives and then the method could mitigate that by a larger weight for vulnerabilities
that are found by more than one scanner. Finally, a study involving the combination and
comparison with static analysis would show the strength and weaknesses of both approaches.

\subsection*{Acknowledgements}

I am grateful to Elmar Juergens for helpful suggestions on the manuscript.

\balance

\end{document}